\documentclass[journal,10pt]{IEEEtran}
\usepackage{amsmath,amsfonts}
\usepackage{algorithmic}
\usepackage{array}
\usepackage[caption=false,font=normalsize,labelfont=sf,textfont=sf]{subfig}
\usepackage{textcomp}
\usepackage{stfloats}
\usepackage{url}
\usepackage{verbatim}
\usepackage{graphicx}
\usepackage[inkscapearea=page]{svg}
\usepackage{adjustbox}

\usepackage{cite}

\usepackage{tikz}
\usetikzlibrary{patterns}
\usepackage{xcolor}
\usepackage{standalone}
\usepackage{ml_package}

\newcommand{\naive}{na\"ive }
\newcommand{\Naive}{Na\"ive }

\begin{document}

\title{Guaranteed Dynamic Scheduling of Ultra-Reliable Low-Latency Traffic  via Conformal~Prediction }

\author{Kfir~M.~Cohen, Sangwoo~Park, Osvaldo~Simeone, Petar~Popovski, and~Shlomo~Shamai~(Shitz),
    \thanks{Kfir M. Cohen, Sangwoo Park, and Osvaldo Simeone are with the King’s Communication, Learning, \& Information Processing (KCLIP) lab, Department of Engineering, King’s College London, WC2R 2LS London, U.K. (e-mail: kfir.cohen@kcl.ac.uk; sangwoo.park@kcl.ac.uk; osvaldo.simeone@kcl.ac.uk).
    Petar Popovski is with the Department of Electronic Systems, Aalborg University, 9220 Aalborg, Denmark (e-mail: petarp@es.aau.dk).
    Shlomo Shamai is with the Viterbi Faculty of Electrical and Computing Engineering, Technion–Israel Institute of Technology, Haifa 3200003, Israel (e-mail: sshlomo@ee.technion.ac.il). The work of KMC, SP, and OS was supported by the European Research Council (ERC) through European Union’s Horizon 2020 Research and Innovation Programme under Grant 725731.
    The work of OS has also been supported by an Open Fellowship of the EPSRC with reference EP/W024101/1.
    The work of PP is supported by the Villum Investigator Grant WATER from the Velux Foundation, Denmark.
    The work of OS and PP was also supported by the European Union’s Horizon Europe project CENTRIC (101096379).
    The work of SS was supported by the European Union’s Horizon 2020 Research And Innovation Programme under Grant 694630. (Corresponding author: Sangwoo Park.)}
    
}
        

\maketitle

\begin{abstract}
The dynamic scheduling of ultra-reliable and low-latency traffic (URLLC) in the uplink can significantly enhance the efficiency of coexisting services, such as enhanced mobile broadband (eMBB) devices, by only allocating resources when necessary. The main challenge is posed by the uncertainty in the process of URLLC packet generation, which mandates the use of predictors for URLLC traffic in the coming frames. In practice, such prediction may overestimate or underestimate the amount of URLLC data to be generated, yielding either an excessive or an insufficient amount of resources to be pre-emptively allocated for URLLC packets. In this paper, we introduce a novel scheduler for URLLC packets that provides formal guarantees on reliability and latency \emph{irrespective of the quality of the URLLC traffic predictor}. The proposed method leverages recent advances in \emph{online conformal prediction (CP)}, and follows the principle of dynamically adjusting the amount of allocated resources so as to meet reliability and latency requirements set by the designer.
\end{abstract}

\vspace{-0.5 cm}
\section{Introduction}
\label{sec: intro}

\begin{figure}[t]
    \centering
    \adjustbox{trim=1.0cm 0.0cm 1.0cm 0.0cm}{
    \includegraphics[trim=1.0cm 0.0cm 1.0cm 0.0cm, clip,width=9.5cm]{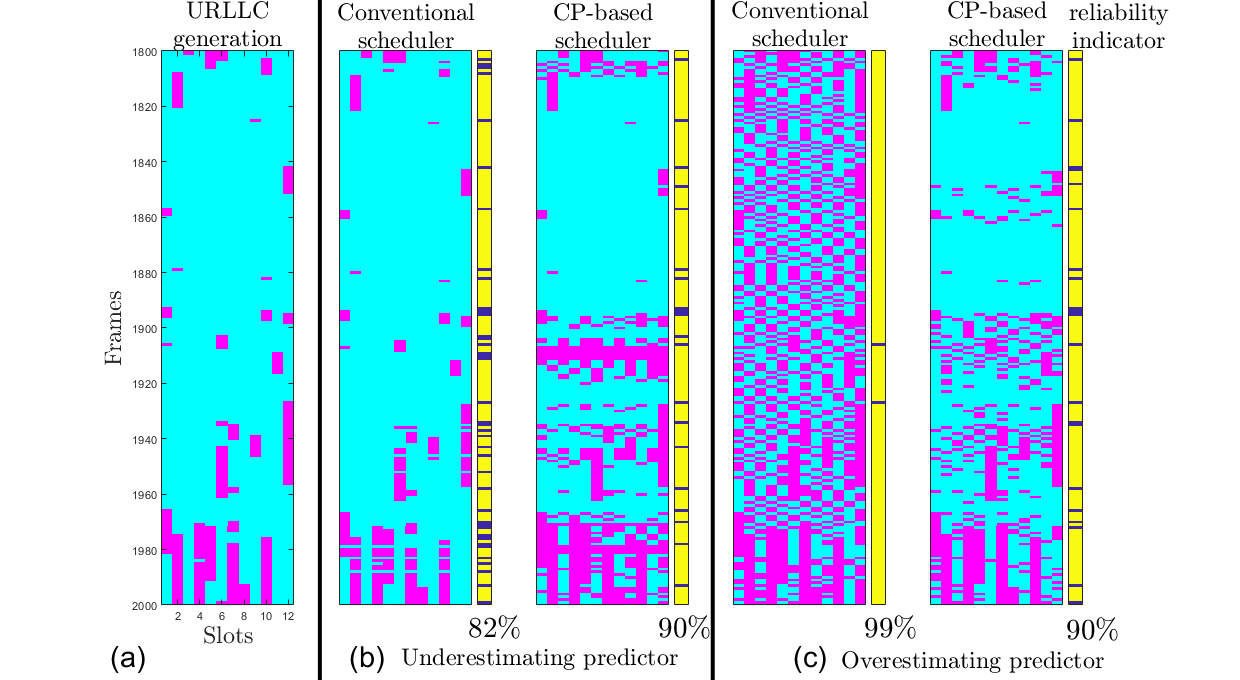}
    }
    \caption{ \textbf{(a)} Data packet generation for URLLC traffic across successive frames (URLLC packets are shown in the darker color). This information is unavailable at the scheduler, which has access only to a predictor that may underestimate or overestimate the number of URLLC packets to be generated (as in parts \textbf{(b)} and \textbf{(c)} respectively). \textbf{(b)} In the former case, a conventional resource allocation scheme that trusts the predictor fails to reliably serve URLLC data (slots allocated for URLLC are in darker color), resulting in an average frame success ratio of $82\%$ that falls short of the target of $90\%$ (for illustrative purposes we set the target unreliability rate to be modest using $\alpha=0.1$, our numerical part uses a tighter value). Scheduling error are shown as darker slots in the sidebar. \textbf{(c)} With an overestimating predictor, a conventional scheduler allocates excessive resources to URLLC traffic, severely impairing eMBB efficiency. eMBB traffic can occupy all slots unassigned to URLLC packets. In either case, the proposed CP-based scheduler is able to meet the URLLC reliability target of $90\%$ by properly adjusting the eMBB spectral efficiency.} \label{fig: fig_allocations}
\end{figure}

\noindent \textbf{\emph{Motivation and overview}}: Servicing \emph{ultra-reliable and low-latency communication} (URLLC) traffic typically calls for a pre-emptive allocation of resources in order to meet stringent delay constraints \cite{cox2020introduction, popovski20185g,vannithamby20205g}. A conservative static allocation of resources for URLLC may guarantee desired levels of reliability and latency, but this comes at the expense of other services, most notably \emph{enhanced mobile broadband (eMBB)}, which cannot use the resources reserved for URLLC. A dynamic allocation of resources, while potentially more efficient, is made challenging by the stochastic nature of URLLC data packet generation, particularly for the uplink \cite{anand2020joint, popovski20185g, kassab2018coexistence, esswie2018opportunistic}. A promising solution is the adoption of predictors of URLLC data packet generation. Concretely, with reference to Fig.~\ref{fig: fig_allocations}, a base station can deploy a predictor of URLLC data packet generation for the following frame, so as to guide the adaptive allocation of slots for URLLC packets, leaving the other slots available for eMBB users.

Such predictors may be based on models that leverage domain knowledge \cite{eggers2019wireless} or statistical information extracted from data \cite{angjelichinoski2019statistical}. In either case, predictions are bound to be imperfect due to model misspecification or to an insufficient access to data \cite{angjelichinoski2019statistical}. Therefore, predictors may consistently overestimate or underestimate the amount of URLLC data to be generated. As a consequence, schedulers that operate on the basis of such predictors would yield either an excessive or an insufficient amount of resources to be pre-emptively allocated for URLLC packets in future frames (see Fig.~\ref{fig: fig_allocations} for an illustration).

In this paper, we introduce a novel scheduler for URLLC packets that provides formal guarantees on reliability and latency \emph{irrespective of the quality of the URLLC traffic predictor}. The proposed method leverages recent advances in \emph{online conformal prediction (CP)} \cite{gibbs2021adaptive,feldman2022achieving}, by dynamically adjusting the amount of allocated resources so as to meet reliability and latency requirements.

\noindent \textbf{\emph{Related work}}: Model-based URLLC traffic predictors, which assume perfect knowledge on the traffic model for optimal allocation strategies, are studied in \cite{alsenwi2019embb, eggers2019wireless, anand2020joint, anand2018resource, ma2020slicing, mahmood2020predictive, abdel2019ultra}. Data-driven approaches \cite{angjelichinoski2019statistical, padilla2021nonlinear, khan2020deep, sun2019learning, zhang2021resource, alsenwi2021intelligent}, which observe data for model training for resource allocation, use tools including unsupervised learning \cite{sun2019learning}, and online learning \cite{zhang2021resource,alsenwi2021intelligent}.

CP is a class of post-hoc calibration methods that transform standard probabilistic model into a \emph{set predictor} that is guaranteed to contain the true target with probability no smaller than a predetermined coverage level \cite{vovk2022algorithmic,shafer2008tutorial}. CP is experiencing a renaissance \cite{barber2022conformal, barber2021predictive, gyorfi2019nearest, park2022few}, with novel applications in \cite{lu2022fair,lu2022three, lindemann2022safe, andeol2023conformal}. \emph{Online CP} alleviates the limitation of conventional CP of requiring a separate calibration data at the cost of providing time-averaged, rather than ensemble, reliability guarantees \cite{gibbs2021adaptive, feldman2022achieving,zaffran2022adaptive,xu2020conformal}. The adoption of CP in communication engineering was proposed in \cite{cohen2022calibrating}, which focused on wireless applications such as symbol demodulation, modulation classification, and received signal strength prediction. 

\noindent \textbf{\emph{Main contributions}}: In this letter, we propose for the first time the application of CP as a design methodology to ensure reliability requirements that hold irrespective of any modeling or data availability assumptions. Specifically, we introduce a CP-based resource allocation scheme for URLLC traffic that makes use of \emph{any} existing model-based or data-driven predictor, offering theoretical reliability guarantees that apply even when the predictor is poorly designed, e.g., due to limited availability of data (see Fig.~\ref{fig: fig_allocations}). The proposed CP-based scheduler is shown via experiments to be capable of efficiently adapting to URLLC traffic, providing eMBB users with a larger fraction of spectral resources as compared to conventional schedulers. Our code is publicly available\footnote{\url{https://github.com/kclip/online_cp_urllc}}.

\begin{figure}[t]
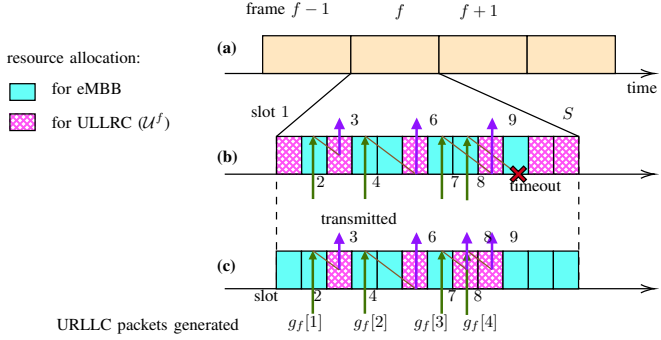

    \centering
    \includestandalone[trim=2.5cm 0cm 0cm 0cm, clip,width=8.9cm]{Figs/fig_tikz_network_slicing}
    \caption{\textbf{(a)} The assumed frame-based communication: each frame $f$ contains $S$ slots that can be allocated for either eMBB  or URLLC traffic. \textbf{(b)} Illustration of the generation of  $G_f=4$ URLLC packets, with each packet generated at a slot marked by an upward arrow incoming into the frame. Each URLLC packet must find an available slot within a maximum delay of $L=2$ slots in order to meet latency requirements. With the given resource allocation, the first three packets are transmitted in the corresponding slots indicated with an upward outgoing arrow, while the fourth packet does not find any available slot within the delay constraint. \textbf{(c)} For the illustrated distinct slot allocation, all URLLC packets are transmitted within the allowed latency of $L=2$ slots.}\label{fig:fig_tikz_network_slicing}
\end{figure}

\vspace{-0.2 cm}
\section{System Model and Problem Definition}\label{sec: System Model}

Fig.~\ref{fig:fig_tikz_network_slicing} illustrates the assumed frame-based transmission setting. Each frame consists of a set $\mathcal{S}=\{1,\dots,S\}$ of $S$ slots, and each of the slots can be allocated either to URLLC or eMBB packets. At the beginning of each frame $f$, a \emph{scheduler} at the base station allocates a subset $\mathcal{U}_f\subseteq\mathcal{S}$ of slots for URLLC transmission, and remaining slots are devoted to eMBB traffic. The main challenge is that the scheduler does not know in advance when URLLC devices will generate packets  \cite{anand2020joint,popovski20185g}.

\noindent \textbf{\emph{URLLC data generation}}: For any frame $f=1,2,\dots$, a total of $G_f \leq S$ URLLC packets are generated. The $i$-th generated packet is produced in the $g_f[i]\in \mathcal{S}$ slot of the frame. As in \cite{kassab2019non} we make the simplifying assumption that no more than one URLLC packet can be generated in a slot. This assumption encodes the requirement that URLLC traffic can be successfully served within any desired degree of reliability by an ideal scheduler that knows the URLLC traffic pattern (or by a trivial scheduler that allocates all slots to URLLC transmissions). The slot indices at which URLLC packets are generated are collected in set $\mathcal{G}_f =\{g_f[1],\dots,g_f[G_f]\}] \subseteq \mathcal{S}$. Importantly, no further assumptions are made on the URLLC data generation mechanism. 

\noindent \textbf{\emph{URLLC latency and reliability constraints}}:  The goal of the scheduler is to allocate the smallest number $U_f=|\mathcal{U}_f|$ of slots, while ensuring that URLLC traffic is served with a prescribed level of latency and reliability. Note that the proposed approach is in line with 3GPP's preemptive scheduling of URLLC traffic on top of eMBB transmissions \cite{cavallero2023}. Specifically, \emph{latency constraints} impose that an URLLC packet generated in time slot $s \in \mathcal{S}$ must be allocated a time slot in the interval $[s,s+1,\dots,\min\{s+L,S\}]$ given maximum allowed latency of $L$ slots. \emph{Reliability} is measured by the fraction of frames $f$ in which \emph{all} $G_f$ URLLC packets are allocated a slot within the described latency constraint of $L$ slots. In particular, we impose that the fraction of frames satisfying this condition is at least $1-\alpha$, for some \emph{unreliability rate} $\alpha \in (0,1)$.
 
To formalize the outlined latency and reliability constraints, we introduce the following definition. We say that a subset $\mathcal{U}_f$ of allocated slots in frame $f$ \emph{``$L$-covers''} a subset $\mathcal{G}_f$ of slots at which URLLC packets are generated if the following condition is met: For each generated URLLC packet $g\in\mathcal{G}_f$, there is a \emph{distinct} allocated URLLC slot $u\in\mathcal{U}_f$ within the latency constraint $L$, i.e., such that the inequalities $0 \leq u-g \leq L$ are satisfied. Note that this condition implies that  the number of allocated slots is no smaller than the number of generated packets, i.e.,  $|\mathcal{U}_f|\geq|\mathcal{G}_f|$.

As an example, in Fig.~\ref{fig:fig_tikz_network_slicing}(b) the allocation  $\mathcal{U}_f~=~\{1,3,6,9,11,12\}$ fails to ``$2$-cover'' the generated set $\mathcal{G}_f=\{2,4,7,8\}$ since the packet generated at $g_f[4]=8$ cannot be served within the latency constraint $L=2$. The allocated slot $9$ ``covers'' the packet generated at $g_f[3]=7$ and hence is unavailable for $g_f[4]=8$, while the remaining allocated slots $11$ and $12$ do not meet the latency constraint. In contrast, the URLLC allocation in Fig.~\ref{fig:fig_tikz_network_slicing}(c) succeeds in $2$-covering the same generated packet $\mathcal{G}_f$.

Given the set of generated packets $\mathcal{G}_f$ and the set of URLLC allocated sets $\mathcal{U}_f$, the reliability measure for frame $f$ is set as the indicator
\vspace{-0.1cm}
\begin{equation}
    r(\mathcal{U}_f|\mathcal{G}_f) = 
     \begin{cases}
         1 & \text{if $\mathcal{U}_f$ $L$-covers $\mathcal{G}_f$}\\
         0 & \text{otherwise.}
     \end{cases} \label{eq:reliability_indicator}
\vspace{-0.1cm}
\end{equation}
Accordingly, given the sequence $\mathcal{U}_{1:F} = \{\mathcal{U}_1,\dots,\mathcal{U}_F\}$ of scheduled slots and the sequence of generated packets $\mathcal{G}_{1:F} = \{\mathcal{G}_1,\dots,\mathcal{G}_F\}$, the \emph{URLLC reliability rate} over a window of $F$ frames is the average reliability measure
\vspace{-0.2cm}
\begin{equation}
    \rho_\text{U} \big(\mathcal{U}_{1:F} \big| \mathcal{G}_{1:F} \big) = \tfrac{1}{F} \sum_{f=1}^{F} r(\mathcal{U}_f | \mathcal{G}_f). \label{eq: reliability empirical}
\vspace{-0.2cm}
\end{equation}
The allocation $\mathcal{U}_{1:F}$ is said to be \emph{$(1-\alpha)$-URLLC reliable} for the generation sequence $\mathcal{G}_{1:F}$ if the following limit holds
\begin{equation}
    \lim_{F\to\infty} \rho_\text{U}\big(\mathcal{U}_{1:F} \big| \mathcal{G}_{1:F} \big) \geq 1-\alpha. \label{eq:reliability_condition}
\end{equation}
This imposes that over a sufficiently long time horizon, the fraction of frames which URLLC packets are served in a timely manner is at least $1-\alpha$.

\noindent \textbf{\emph{eMBB efficiency}}: A scheduler could easily obtain the highest coverage rate of $1$ by allocating all $S$ slots to URLLC traffic. However, this would come at the cost of eMBB traffic. The \emph{eMBB efficiency} of an allocation strategy is measured by the fraction of slots available for eMBB transmission over a window of $F$ frames, i.e., as
\vspace{-0.3cm}
\begin{equation}
    \eta_\text{e} (\mathcal{U}_{1:F}) = \tfrac{1}{F} \sum_{f=1}^F \tfrac{S-|\mathcal{U}_f|}{S} = 1 - \tfrac{1}{F S} \sum_{f=1}^F |\mathcal{U}_f| .
    \label{eq:eMBB_efficiency}
\vspace{-0.2cm}
\end{equation}
Since the reliability requirements of URLLC are more stringent, by many orders of magnitude, as compared to eMBB, we focus on meeting URLLC reliability constraints, while serving eMBB traffic is in a best-effort fashion.

\noindent \textbf{\emph{URLLC predictor}}:  The scheduler has access to an arbitrary probabilistic URLLC traffic predictor. The predictor may be model-based, e.g., based on a Markov model, or data-driven, e.g., a recurrent neural network, and we make no assumptions on its accuracy. The predictor outputs a probability distribution $q_f(\cdot)$ over all possible subsets of the slot set $\mathcal{S}$. Accordingly, the predictor assigns a probability $q_f(\mathcal{G}_f)$ to each subset $\mathcal{G}_f$ of possible slot indices containing URLLC packets in frame $f$. This probability generally depends in arbitrary ways on the past observations of the predictor. Such observations include the past decisions $\mathcal{U}_{1:f-1}$ of the scheduler, as well as, possibly partial, information about the previous packet generation subsets $\mathcal{G}_{1:f-1}$. For instance, the predictor may have access to the previous reliability indicators $r(\mathcal{U}_{f^\prime}|\mathcal{G}_{f^\prime})$ with $f^\prime=1,...,f-1$ providing information about whether past allocations have been successful or not. Furthermore, while the probability $q_f(\cdot)$ generally ranges over all possible $2^S$ subsets of slots, practical predictors may, e.g., factorize this distribution so as to reduce complexity \cite{cammerer2017scaling, shlezinger2020deepsic}.

\vspace{-0.2 cm}
\section{CP-Based URLLC Resource Allocation}\label{sec: CP-Based URLLC Resource Allocation}

In this section, we introduce the proposed CP-based resource scheduler, proven to satisfy the reliability constraint \eqref{eq:reliability_condition} irrespective of the quality of the predictor $q_f(\cdot)$ on which its decisions are based. This important result is obtained by suitably adjusting the number of slots allocated to URLLC traffic, and hence the resulting eMBB efficiency \eqref{eq:eMBB_efficiency}. We start by reviewing a \naive approach to scheduling that ``trusts'' the predictor to be accurate and well-calibrated.
\vspace{-0.4cm}

 \subsection{\Naive Prediction-Based Scheduler}\label{sec: Naive Prediction-Based Scheduler}
 \vspace{-0.1cm}
 
 Assume that the predictor $q_f(\cdot)$ is well-calibrated, in the sense that it provides the actual probability $q_f(\mathcal{G}_f)$ that a certain URLLC traffic pattern $\mathcal{G}_f$ is realized. For model-based predictors, this would be the case if the available domain knowledge is extremely precise; and for data-driven predictors this condition may arise if one has access to large amount of relevant data. Under such ideal conditions, a \naive scheduler would aim at minimizing the number $|\mathcal{U}_f|$ of allocated slots under the constraint that the sum of probabilities $q_f(\mathcal{G})$ across all arrivals 
 $\mathcal{G}$ that are $L$-covered by $\mathcal{U}_f$ is no smaller than $1-\alpha$. We propose to address this combinatorial problem through a two-step heuristic approach. First, we find the smallest set $\mathbf{\Gamma}$ of slot generation patterns $\mathcal{G}_f$ to which the predictor $q_f(\cdot)$ assigns a probability at least $1-\alpha$, i.e., we first solve the problem
\vspace{-0.1cm}
\begin{equation}
    \mathbf{\Gamma} (\alpha|q_f)=  \underset{\mathbf{\Gamma} \in 2^S}{\argmin}\: \quad |\mathbf{\Gamma}| \quad
    \text{s.t.} \quad \sum_{\mathcal{G} \in \mathbf{\Gamma}} q_f(\mathcal{G}) \geq 1-\alpha.  \label{eq: Gamma NPB} 
    \vspace{-0.2 cm}
\end{equation}
This problem can be addressed by sorting the probabilities $q_f(\cdot)$ in decreasing order. Note that, in practice, problem \eqref{eq: Gamma NPB} can be simplified by restricting the domain, e.g., by considering only traffic patterns of no more than $G_{\text{max}}$ packets.

Once a set $\mathbf{\Gamma} (\alpha|q_f)$ of subsets is identified, the scheduler could find an allocation $\mathcal{U}_f$ that guarantees that, for all patterns $\mathcal{G}_f \in \mathbf{\Gamma} (\alpha|q_f)$, we have $r(\mathcal{U}_f|\mathcal{G}_f)=1$ and hence all URLLC packets are correctly transmitted within the latency condition. A greedy algorithm satisfying this condition is detailed in Algorithm~\ref{alg:greedy_allocation}. The approach operates backwards from slot $S$ to slot $1$. For any slot $s$ that belongs to any of the traffic patterns in set $\mathbf{\Gamma}$, the slot $s$ is added to the set of allocated slots $\mathcal{U}$. Furthermore, for each pattern $\mathcal{G} \in \mathbf{\Gamma}$, one slot $s'\leq s$ is removed if it is the largest not yet considered and if it is within $L$ time slots of the allocated slot $s$.

Under suitable ergodicity conditions (see, e.g., \cite{gray2009probability}), making the strong assumption that the predictor is indeed well-accurate, the reliability inequality \eqref{eq:reliability_condition} would be satisfied by the \naive scheduler with probability 1. 

\begin{algorithm}[t] 
    \DontPrintSemicolon
    \smallskip
    \KwIn{latency constraint $L$, set of subsets $\mathbf{\Gamma}$}
    \KwOut{URLLC slot allocation $\mathcal{U}$}
    \vspace{0.15cm}
    \hrule
    \vspace{0.15cm}
    {\bf initialize} slot allocation $\mathcal{U}=\emptyset$ \\
    \For{\em $s=S,S-1,\ldots,1$}{
        \If{$s \in \cup_{\mathcal{G} \in \mathbf{\Gamma}}\mathcal{G}$}{
            $\mathcal{U} \leftarrow \mathcal{U} \cup \{ s \} $  
            \For{{\em $\mathcal{G}\in\mathbf{\Gamma}$ }}{
                $\mathcal{G} \gets \mathcal{G} \setminus \big\{ \max \big( \{s-L,\dots,s\}\cap \mathcal{G} \big) \big\}$
            }
        }
    }
    \KwRet $\mathcal{U}$
    \caption{Greedy Slot Allocation}
    \label{alg:greedy_allocation}
\end{algorithm}

\vspace{-0.3cm}
\subsection{CP-Based Scheduler}\label{sec: CP-Based Scheduler}

In practice, one cannot rely on the accuracy of the predictor to guarantee the  reliability condition \eqref{eq:reliability_condition}. Inspired by online CP \cite{gibbs2021adaptive, feldman2022achieving}, we now introduce an approach that is guaranteed to meet the condition \eqref{eq:reliability_condition} no matter what the accuracy of the predictor is and for every realization of URLLC traffic patterns. While not affecting URLLC reliability, the accuracy of the predictor dictates eMBB efficiency \eqref{eq:eMBB_efficiency}, with a more accurate predictor yielding a higher eMBB efficiency. 

The key idea is to adjust the threshold used in the definition of set \eqref{eq: Gamma NPB} as a function of the past reliability measures, so as to meet the reliability condition \eqref{eq:reliability_condition}. Let us define as $\alpha_f$ the target unreliability rate for frame $f$, which is used in \eqref{eq: Gamma NPB} to obtain the set $ \mathbf{\Gamma}(\alpha_f|q_f)$. A smaller value of $\alpha_f$ yields a larger set $ \mathbf{\Gamma}(\alpha_f|q_f)$. Once such a set is identified, the CP-based scheduler applies the same greedy approach as the \naive scheme to identify set $\mathcal{U}_f$ (see Algorithm~\ref{alg:greedy_allocation}). Intuitively, the target unreliability rate $\alpha_{f+1}$ for frame $f+1$ should be chosen to be small when the average success rate $f^{-1} \sum_{f^\prime=1}^{f} r(\mathcal{U}_{f^\prime}|\mathcal{G}_{f^\prime})$ obtained so far is smaller than $1-\alpha$; and one should increase $\alpha_{f+1}$ if the average success rate so far is larger than $1-\alpha$.

To this end, we assume that at the end of the $f$-th frame the scheduler gains access to the reliability measure $r (\mathcal{U}_f | \mathcal{G}_f ) $. In practice, this requires some minimal feedback from URLLC devices informing the base station of an unsuccessful attempt to transmit a packet. Then, the target per-frame unreliability threshold $\alpha_{f+1}$ is set as  $\alpha_{f+1}=\varphi(\theta_{f+1})$, where $\varphi(\cdot)$ is a monotonically increasing function, known as the \emph{stretching function} \cite{feldman2022achieving}. The parameter $\theta_{f+1}$ is updated as
\begin{equation}
    \theta_{f+1} \gets \theta_f + \gamma \big( r (\mathcal{U}_f | \mathcal{G}_f ) - (1-\alpha)\big),\label{eq:theta_f update}
\end{equation}
where $\gamma>0$ is an update step. We adopt the stretching function
\begin{equation}
    \varphi(\theta) = \tfrac{1}{2} \Big( 1+\sin\big(\pi\big(\max\big\{0,\min\{1,\theta\}\big\}-0.5\big)\big) \Big),  \label{eq: stretching func quantile based}
\end{equation} 
which satisfies the conditions in \cite[Theorem 1]{feldman2022achieving}. 

By \cite[Proposition 4.1]{gibbs2021adaptive}, this choice ensures that the difference between the URLLC reliability rate, $\rho_\text{U}(\mathcal{U}_{1:F}|\mathcal{G}_{1:F})$, and the target rate $1-\alpha$ satisfies the inequality \begin{equation}\big|\rho_\text{U}(\mathcal{U}_{1:F}|\mathcal{G}_{1:F})-(1-\alpha)\big|\leq \mathcal{O}(1/F)\end{equation} for any number of frames, $F$, and irrespective of the specific realized sequence of traffic patterns. This condition  yields the limit \eqref{eq:reliability_condition} as the number of frames, $F$, grows large. 

\begin{algorithm}[t] 
    \DontPrintSemicolon
    \smallskip
    \KwIn{target unreliability rate $\alpha>0$, probabilistic predictor $\{q_f\}_{f\in\mathbb{N}}$,  latency constraint $L$, update step $\gamma>0$}
    \KwOut{URLLC slot allocations $\mathcal{U}_1,\mathcal{U}_2,\dots$}
    \vspace{0.15cm}
    \hrule
    \vspace{0.15cm}
    {\bf initialize} threshold $\theta_1\gets\varphi^{-1}(\alpha)$ \\
    \For{{\em $f=1,\ldots,$}}{
        find $\mathcal{U}_f$ using Algorithm~\ref{alg:greedy_allocation} for $\mathbf{\Gamma}(\varphi(\theta_f)|q_f)$ using \eqref{eq: Gamma NPB}\\
        allocate slots $\mathcal{U}_f$ for URLLC traffic $\mathcal{G}_f$ in frame $f$\\
        obtain reliability indicator $r(\mathcal{U}_f | \mathcal{G}_f )$ via \eqref{eq:reliability_indicator}\\
        $\theta_{f+1} \gets \theta_f + \gamma\big(r(\mathcal{U}_f | \mathcal{G}_f ) - (1-\alpha)\big)$\\
        }
    \KwRet $\mathcal{U}_1,\mathcal{U}_2,\dots$\\
    \caption{CP-Based Scheduler}
    \label{alg:overall}
\end{algorithm}

\vspace{-0.2 cm}
\section{Experiments and Conclusions}\label{sec: Experiments and Conclusions}

To validate the proposed approach, we conducted experiments under a Markov packet generation mechanism. Recall that the proposed scheme provides guarantees that do not depend on the statistics of the packet arrival process. The arrival process is defined by four parameters $(p^-,p^+,G_{\text{min}},G_{\text{max}})$. Accordingly, given the current traffic pattern $\mathcal{G}_f$, the next traffic pattern  $\mathcal{G}_{f+1}$ has a number of packets equal to $G_{f+1}=[G_f+W_{f+1}]_{G_{\text{min}}}^{G_{\text{max}}}$, where $W_f$ is a ternary variable that equals $W_{f+1}=1$ with probability $p^+$, $W_{f+1}=-1$ with probability $p^-$, and  $W_{f+1}=0$ otherwise. The function $[\cdot]_{G_{\text{min}}}^{G_{\text{max}}}$ clips the input argument within the range $[G_{\text{min}},G_{\text{max}}]$. Given a number $G_{f+1}\neq G_f$ of packets, the traffic pattern $\mathcal{G}_{f+1}$ is selected uniformly among all subsets of cardinality $G_{f+1}$ that can be obtained from pattern $\mathcal{G}_{f}$ by adding a slot (if $G_{f+1}>G_f$) or removing a slot (if $G_{f+1}<G_f$). Otherwise, if $G_{f+1}=G_f$, we set $\mathcal{G}_{f+1}=\mathcal{G}_{f}$.  While simplistic, this mechanism allows us to draw insightful conclusions on the role of predictors in the performance of schedulers.  

To this end, we assume that the predictor $q_f(\cdot)$ adopts the same Markov model of the ground-truth packet generation mechanism, but with generally mismatched probabilities $\hat{p}^+$ and $ \hat{p}^-$ in lieu of the true probabilities $p^+$ and $ p^-$.

Fig.~\ref{fig: fig_allocations} shows the generated packets $\{\mathcal{G}_f\}$ over the last $200$ frames of a $2000$ frames run, along with the allocation $\{\mathcal{U}_f\}$ and reliability indicators \eqref{eq:reliability_indicator} in the side bars. Each frame consists of $S=12$ slots, the URLLC latency is $L=1$, the learning rate $\gamma=0.1$, and traffic follows $G^\text{min}=0$ and $G^\text{max}=6$ and $p^+ = p^- = 0.16$. We consider two predictors: The first underestimates the parameters with $\hat{p}^+ = \hat{p}^- = 0.02$, while the second overestimates $\hat{p}^+ = \hat{p}^- = 0.40$. The conventional scheduler either fails to meet \eqref{eq:reliability_condition} using the underestimating predictor (covering $82\%$ instead of $1-\alpha=90\%$), or allocates an excessively large number of slots using the overestimating predictor. In contrast, the CP-based predictor can effectively adjust the eMBB efficiency to the quality of the predictor, always meeting the reliability constraint \eqref{eq:reliability_condition}. For example, it trades excessive coverage ($98\%$ to $90\%$) into higher eMBB efficiency ($45\%$ to $66\%$ as in Fig.~\ref{fig: fig_allocations}(c)).

\begin{figure}[t]
    \centering
    \includegraphics[trim=1.2cm 0cm 2.5cm 0.2cm, clip,width=9cm]{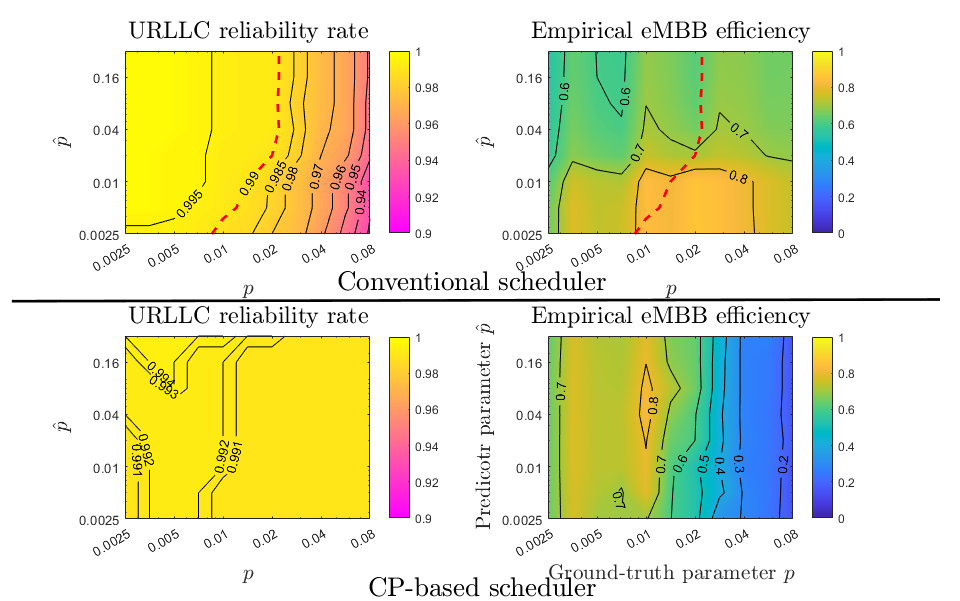}
    \vspace{-0.3cm}
    \caption{URLLC reliability rate \eqref{eq: reliability empirical} and eMBB efficiency \eqref{eq:eMBB_efficiency} for conventional scheduler (Sec.~\ref{sec: Naive Prediction-Based Scheduler}) and CP-based scheduler (Sec.~\ref{sec: CP-Based Scheduler}) as a function of the ground-truth traffic parameter and predictor parameter. The target rate is $1-\alpha=0.99$ (dashed red line for conventional scheduler; the CP-based scheduler always satisfies the reliability condition).
    }\label{fig: fig_coverage_effic}
\end{figure}

We now set $\alpha=0.01$ and $\gamma=0.05$, and investigate the impact of a mismatch between the URLLC traffic model assumed by the predictor and the ground-truth model. We set $p^+ = p^- = p$ and $\hat{p}^+ = \hat{p}^- = \hat{p}$, and let both parameters vary. Fig.~\ref{fig: fig_coverage_effic} shows the empirical URLLC reliability rate \eqref{eq: reliability empirical} and the empirical eMBB efficiency \eqref{eq:eMBB_efficiency} at the completion of $F=4000$ frames for both the \naive scheduler and the CP-based scheduler. The \naive scheduler is significantly affected by a mismatch between predictor and ground-truth packet generation mechanism, yielding either ill empirical coverage (below $1-\alpha=0.99$) or over coverage. In contrast, the CP-based predictor is able to flatten the coverage to asymptotically reach the long-term target $1-\alpha$.

\newpage

\bibliographystyle{IEEEtran} 
\bibliography{my_bib.bib} 

\begin{thebibliography}{10}
\providecommand{\url}[1]{#1}
\csname url@samestyle\endcsname
\providecommand{\newblock}{\relax}
\providecommand{\bibinfo}[2]{#2}
\providecommand{\BIBentrySTDinterwordspacing}{\spaceskip=0pt\relax}
\providecommand{\BIBentryALTinterwordstretchfactor}{4}
\providecommand{\BIBentryALTinterwordspacing}{\spaceskip=\fontdimen2\font plus
\BIBentryALTinterwordstretchfactor\fontdimen3\font minus
  \fontdimen4\font\relax}
\providecommand{\BIBforeignlanguage}[2]{{%
\expandafter\ifx\csname l@#1\endcsname\relax
\typeout{** WARNING: IEEEtran.bst: No hyphenation pattern has been}%
\typeout{** loaded for the language `#1'. Using the pattern for}%
\typeout{** the default language instead.}%
\else
\language=\csname l@#1\endcsname
\fi
#2}}
\providecommand{\BIBdecl}{\relax}
\BIBdecl

\bibitem{cox2020introduction}
C.~Cox, \emph{{An Introduction to {5G}: The New Radio, {5G} Network and
  Beyond}}.\hskip 1em plus 0.5em minus 0.4em\relax John {W}iley \& {S}ons,
  2020.

\bibitem{popovski20185g}
P.~Popovski, K.~F. Trillingsgaard, O.~Simeone, and G.~Durisi, ``{5G Wireless
  Network Slicing for eMBB, URLLC, and mMTC: A Communication-Theoretic View},''
  \emph{IEEE Access}, vol.~6, pp. 55\,765--55\,779, 2018.

\bibitem{vannithamby20205g}
R.~Vannithamby and A.~Soong, \emph{{5G Verticals: Customizing Applications,
  Technologies and Deployment Techniques}}.\hskip 1em plus 0.5em minus
  0.4em\relax John Wiley \& Sons, 2020.

\bibitem{anand2020joint}
A.~Anand, G.~De~Veciana, and S.~Shakkottai, ``{Joint Scheduling of URLLC and
  eMBB Traffic in 5G Wireless Networks},'' \emph{IEEE/ACM Transactions on
  Networking}, vol.~28, no.~2, pp. 477--490, 2020.

\bibitem{kassab2018coexistence}
R.~Kassab, O.~Simeone, and P.~Popovski, ``{Coexistence of URLLC and eMBB
  Services in the C-RAN Uplink: An Information-Theoretic Study},'' in
  \emph{2018 IEEE Global Communications Conference (GLOBECOM)}.\hskip 1em plus
  0.5em minus 0.4em\relax IEEE, 2018, pp. 1--6.

\bibitem{esswie2018opportunistic}
A.~A. Esswie and K.~I. Pedersen, ``{Opportunistic Spatial Preemptive Scheduling
  for URLLC and eMBB Coexistence in Multi-User 5G Networks},'' \emph{Ieee
  Access}, vol.~6, pp. 38\,451--38\,463, 2018.

\bibitem{eggers2019wireless}
P.~C. Eggers, M.~Angjelichinoski, and P.~Popovski, ``Wireless channel modeling
  perspectives for ultra-reliable communications,'' \emph{IEEE Transactions on
  Wireless Communications}, vol.~18, no.~4, pp. 2229--2243, 2019.

\bibitem{angjelichinoski2019statistical}
M.~Angjelichinoski, K.~F. Trillingsgaard, and P.~Popovski, ``{A Statistical
  Learning Approach to Ultra-Reliable Low Latency Communication},'' \emph{IEEE
  Transactions on Communications}, vol.~67, no.~7, pp. 5153--5166, 2019.

\bibitem{gibbs2021adaptive}
\BIBentryALTinterwordspacing
I.~Gibbs and E.~Candès, ``{Adaptive Conformal Inference Under Distribution
  Shift},'' 2021. [Online]. Available: \url{https://arxiv.org/abs/2106.00170}
\BIBentrySTDinterwordspacing

\bibitem{feldman2022achieving}
\BIBentryALTinterwordspacing
S.~Feldman, L.~Ringel, S.~Bates, and Y.~Romano, ``{Achieving Risk Control in
  Online Learning Settings},'' 2022. [Online]. Available:
  \url{https://arxiv.org/abs/2205.09095}
\BIBentrySTDinterwordspacing

\bibitem{alsenwi2019embb}
M.~Alsenwi, N.~H. Tran, M.~Bennis, A.~K. Bairagi, and C.~S. Hong, ``{eMBB-URLLC
  Resource Slicing: A Risk-Sensitive Approach},'' \emph{IEEE Communications
  Letters}, vol.~23, no.~4, pp. 740--743, 2019.

\bibitem{anand2018resource}
A.~Anand and G.~de~Veciana, ``{Resource Allocation and HARQ Optimization for
  URLLC Traffic in 5G Wireless Networks},'' \emph{IEEE Journal on Selected
  Areas in Communications}, vol.~36, no.~11, pp. 2411--2421, 2018.

\bibitem{ma2020slicing}
T.~Ma, Y.~Zhang, F.~Wang, D.~Wang, and D.~Guo, ``{Slicing Resource Allocation
  for eMBB and URLLC in 5G RAN},'' \emph{Wireless Communications and Mobile
  Computing}, vol. 2020, pp. 1--11, 2020.

\bibitem{mahmood2020predictive}
N.~H. Mahmood, O.~A. Lopez, H.~Alves, and M.~Latva-Aho, ``{A Predictive
  Interference Management Algorithm for URLLC in Beyond 5G Networks},''
  \emph{IEEE Communications Letters}, vol.~25, no.~3, pp. 995--999, 2020.

\bibitem{abdel2019ultra}
M.~K. Abdel-Aziz, S.~Samarakoon, M.~Bennis, and W.~Saad, ``{Ultra-Reliable and
  Low-Latency Vehicular Communication: An Active Learning Approach},''
  \emph{IEEE Communications Letters}, vol.~24, no.~2, pp. 367--370, 2019.

\bibitem{padilla2021nonlinear}
C.~Padilla, R.~Hashemi, N.~H. Mahmood, and M.~Latva-Aho, ``{A Nonlinear
  Autoregressive Neural Network for Interference Prediction and Resource
  Allocation in URLLC Scenarios},'' in \emph{2021 International Conference on
  Information and Communication Technology Convergence (ICTC)}, 2021, pp.
  184--189.

\bibitem{khan2020deep}
H.~Khan, M.~M. Butt, S.~Samarakoon, P.~Sehier, and M.~Bennis, ``{Deep Learning
  Assisted CSI Estimation for Joint URLLC and eMBB Resource Allocation},'' in
  \emph{2020 IEEE International Conference on Communications Workshops (ICC
  Workshops)}, 2020, pp. 1--6.

\bibitem{sun2019learning}
C.~Sun and C.~Yang, ``{Learning to Optimize with Unsupervised Learning:
  Training Deep Neural Networks for URLLC},'' in \emph{2019 IEEE 30th Annual
  International Symposium on Personal, Indoor and Mobile Radio Communications
  (PIMRC)}, 2019, pp. 1--7.

\bibitem{zhang2021resource}
J.~Zhang, C.~Sun, and C.~Yang, ``{Resource Allocation in URLLC with Online
  Learning for Mobile Users},'' in \emph{2021 IEEE 93rd Vehicular Technology
  Conference (VTC2021-Spring)}, 2021, pp. 1--5.

\bibitem{alsenwi2021intelligent}
M.~Alsenwi, N.~H. Tran, M.~Bennis, S.~R. Pandey, A.~K. Bairagi, and C.~S. Hong,
  ``{Intelligent Resource Slicing for eMBB and URLLC Coexistence in 5G and
  Beyond: A Deep Reinforcement Learning Based Approach},'' \emph{IEEE
  Transactions on Wireless Communications}, vol.~20, no.~7, pp. 4585--4600,
  2021.

\bibitem{vovk2022algorithmic}
V.~Vovk, A.~Gammerman, and G.~Shafer, \emph{{Algorithmic Learning in a Random
  World}}.\hskip 1em plus 0.5em minus 0.4em\relax Springer Nature, 2022.

\bibitem{shafer2008tutorial}
G.~Shafer and V.~Vovk, ``{A Tutorial on Conformal Prediction},'' \emph{Journal
  of Machine Learning Research}, vol.~9, no.~3, 2008.

\bibitem{barber2022conformal}
R.~F. Barber, E.~J. Candes, A.~Ramdas, and R.~J. Tibshirani, ``{Conformal
  Prediction Beyond Exchangeability},'' \emph{arXiv preprint arXiv:2202.13415},
  2022.

\bibitem{barber2021predictive}
------, ``{Predictive Inference with the Jackknife+},'' \emph{The Annals of
  Statistics}, vol.~49, no.~1, pp. 486--507, 2021.

\bibitem{gyorfi2019nearest}
L.~Gy{\^o}rfi and H.~Walk, ``{Nearest Neighbor Based Conformal Prediction},''
  in \emph{Annales de l'ISUP}, vol.~63, no. 2-3, 2019, pp. 173--190.

\bibitem{park2022few}
S.~Park, K.~M. Cohen, and O.~Simeone, ``{Few-Shot Calibration of Set Predictors
  via Meta-Learned Cross-Validation-Based Conformal Prediction},'' \emph{arXiv
  preprint arXiv:2210.03067}, 2022.

\bibitem{lu2022fair}
C.~Lu, A.~Lemay, K.~Chang, K.~Höbel, and J.~Kalpathy-Cramer, ``{Fair Conformal
  Predictors for Applications in Medical Imaging},'' in \emph{Proceedings of
  the AAAI Conference on Artificial Intelligence}, vol.~36, no.~11.\hskip 1em
  plus 0.5em minus 0.4em\relax PMLR, 2022, pp. 12\,008--12\,016.

\bibitem{lu2022three}
C.~Lu, K.~Chang, P.~Singh, and J.~Kalpathy-Cramer, ``{Three Applications of
  Conformal Prediction for Rating Breast Density in Mammography},'' \emph{arXiv
  preprint arXiv:2206.12008}, 2022.

\bibitem{lindemann2022safe}
L.~Lindemann, M.~Cleaveland, G.~Shim, and G.~J. Pappas, ``{Safe Planning in
  Dynamic Environments using Conformal Prediction},'' \emph{arXiv preprint
  arXiv:2210.10254}, 2022.

\bibitem{andeol2023conformal}
\BIBentryALTinterwordspacing
L.~Andéol, T.~Fel, F.~De~Grancey, and L.~Mossina, ``{Conformal Prediction for
  Trustworthy Detection of Railway Signals},'' 2023. [Online]. Available:
  \url{https://arxiv.org/abs/2301.11136}
\BIBentrySTDinterwordspacing

\bibitem{zaffran2022adaptive}
M.~Zaffran, O.~F{\'e}ron, Y.~Goude, J.~Josse, and A.~Dieuleveut, ``{Adaptive
  Conformal Predictions for Time Series},'' in \emph{International Conference
  on Machine Learning}.\hskip 1em plus 0.5em minus 0.4em\relax PMLR, 2022, pp.
  25\,834--25\,866.

\bibitem{xu2020conformal}
C.~Xu and Y.~Xie, ``{Conformal Prediction for Dynamic Time-Series},''
  \emph{arXiv preprint arXiv:2010.09107}, 2020.

\bibitem{cohen2022calibrating}
\BIBentryALTinterwordspacing
K.~M. Cohen, S.~Park, O.~Simeone, and S.~Shamai, ``{Calibrating AI Models for
  Wireless Communications via Conformal Prediction},'' 2022. [Online].
  Available: \url{https://arxiv.org/abs/2212.07775}
\BIBentrySTDinterwordspacing

\bibitem{kassab2019non}
R.~Kassab, O.~Simeone, P.~Popovski, and T.~Islam, ``{Non-Orthogonal
  Multiplexing of Ultra-Reliable and Broadband Services in Fog-Radio
  Architectures},'' \emph{IEEE Access}, vol.~7, pp. 13\,035--13\,049, 2019.

\bibitem{cavallero2023}
S.~Cavallero, N.~S. Grande, F.~Pase, M.~Giordani, J.~Eichinger, R.~Verdone, and
  M.~Zorzi, ``{A New Scheduler for URLLC in 5G NR IIoT Networks with
  Spatio-Temporal Traffic Correlations},'' in \emph{2017 IEEE International
  Conference on Communications (ICC) in Rome, Italy}.\hskip 1em plus 0.5em
  minus 0.4em\relax IEEE, 2023.

\bibitem{cammerer2017scaling}
S.~Cammerer, T.~Gruber, J.~Hoydis, and S.~Ten~Brink, ``{Scaling Deep
  Learning-Based Decoding of Polar Codes via Partitioning},'' in \emph{GLOBECOM
  2017-2017 IEEE Global Communications Conference}.\hskip 1em plus 0.5em minus
  0.4em\relax IEEE, 2017, pp. 1--6.

\bibitem{shlezinger2020deepsic}
N.~Shlezinger, R.~Fu, and Y.~C. Eldar, ``{DeepSIC: Deep Soft Interference
  Cancellation for Multiuser MIMO Detection},'' \emph{IEEE Transactions on
  Wireless Communications}, vol.~20, no.~2, pp. 1349--1362, 2020.

\bibitem{gray2009probability}
R.~M. Gray and R.~Gray, \emph{{Probability, Random Processes, and Ergodic
  Properties}}.\hskip 1em plus 0.5em minus 0.4em\relax Springer, 2009, vol.~1.

\end{thebibliography}

\vspace{11pt}

\vfill

\end{document}